\documentclass[prb,aps,amsmath,amssymb,reprint,showpacs,floatfix,superscriptaddress]{revtex4-1}
\usepackage{graphicx}
\usepackage{dcolumn}
\usepackage{bm}
\usepackage{mathtools}

\begin{document}



\title{Majorana states and magnetic orbital motion in planar hybrid nanowires}


\author{Javier Osca}
\email{javier@ifisc.uib-csic.es}
\affiliation{Institut de F\'{\i}sica Interdisciplin\`aria i de Sistemes Complexos IFISC (CSIC-UIB), E-07122 Palma de Mallorca, Spain}
\author{Lloren\c{c} Serra}
\affiliation{Institut de F\'{\i}sica Interdisciplin\`aria i de Sistemes Complexos IFISC (CSIC-UIB), E-07122 Palma de Mallorca, Spain}
\affiliation{Departament de F\'{\i}sica, Universitat de les Illes Balears, E-07122 Palma de Mallorca, Spain}

\date{May 27, 2015}

\begin{abstract}
The Majorana phase boundaries in planar 2D hybrid (semiconductor-superconductor) nanowires are modified by orbital effects
due to off plane magnetic components. We show that Majorana zero modes survive sizable  vertical field tiltings, 
uncovering a remarkable phase diagram. Analytical expressions of the phase boundaries are given for the strong orbital limit. 
These phase boundaries can be fulfilled with attainable setups, such as an InAs nanowire of $150\, {\rm nm}$ in transverse width.
\end{abstract}


\maketitle 

\section{Introduction}
The physics of 2D electron gases in magnetic fields has proved invaluable for the  Condensed Matter field with, e.g., the celebrated quantum Hall effect \cite{cha} as well as with many devices based on quantum wells, wires and dots.\cite{ihn} On the other hand, Majorana zero modes in quasi 1D wires  have recently  attracted strong interest, both as exotic quasi-particles and as candidates for topological quantum computing.\cite{Alicea,Beenakker,StanescuREV,Franz}
In this article we relate 2D-gas properties and Majorana physics, showing the remarkable role of the
orbital motions characteristic of 2D systems in magnetic fields. 

Majorana modes in quasi 1D wires are effectively chargeless, zero-energy quasiparticles. They arise  from the splitting, through a phase transition, of bulk electronic states into pairs of quasiparticles on the wire ends, each one  being its own 
antiparticle. \cite{Lutchyn,Oreg,Stanescu,Flensberg,Ganga,Lin,Klino2,Stic,Liu,Jiang,Pien,San} 
An important feature of the Majorana mode is that it appears only when a critical value of the external magnetic field, a phase-transition threshold, has been surpassed.
Several  experiments with hybrid superconductor-semiconductor nanowires using tunneling spectroscopy from a normal conductor to the nanowire have observed a zero bias peak consistent with a Majorana 
state.\cite{Mourik,Deng,Das,Finck}
The observed peak height is, however, an order of magnitude lower than the quantized value $2e^2/h$. This discrepancy is not yet well understood, as it might be due to effects ranging from finite temperature, experimental and tunneling resolutions to other low-energy subgap states and possible inelastic and renormalization processes.\cite{Pen,Sar_prep}

In practice, distinguishing zero-bias peaks 
due to Majorana modes
from other potential sources relies on the detailed knowledge 
of the  phase diagram in each particular physical realization. Therefore, it is highly relevant knowing how Majorana physics is affected by the extra dimension in 2D, with respect to 1D.  This question has been
addressed with quasi-1D {\em multiband} wires,\cite{Lutchyn2,Potter,Potter2,sanjose} modelling transverse modes as  mutually coupled 1D wires.
This leads to essentially the 1D physics with only one Majorana mode at each nanowire end 
when an odd number of transverse modes are above their critical magnetic field. However, the role of the magnetic orbital motion has been usually disregarded.  
Addressing Majorana physics in 2D systems with orbital motion is therefore relevant as a way to discard alternative scenarios that have been suggested, such as attributing  the observations to Kondo-like interaction effects.\cite{sanjose}
We want to clarify too that the above mentioned experiments
\cite{Mourik,Deng,Das,Finck} used cylindrical nanowires, for which the orbital effects
may be different from those discussed here.

\begin{figure}[t]
\centering
\resizebox{0.35\textwidth}{!}{
	\includegraphics{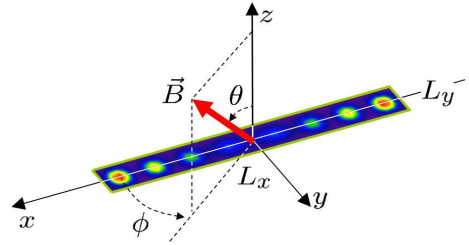}
}
\caption{Schematic of a 2D planar nanowire
showing the axis definitions. A magnetic field in a tilted direction
is included. The density distribution of Majorana modes on the 
wire ends is qualitatively shown.}
\label{F1}
\end{figure}

In this article we show that in a planar nanowire the orbital motion strongly affects  the phase transition boundaries. We show how the Majorana phase survives sizable vertical field  tiltings (Fig.\ \ref{F1}), even reaching the purely perpendicular orientation in some cases. 
This is not an intuitive result since the electronic orbital motion might lead to a gap closing, 
allowing edge propagating solutions, or it might totally change the character of the topological states. 
In this sense we note that the s-wave nanowires inside a magnetic field have an associated directionality with their Majoranas located at the left and right edges (Fig.\ \ref{F1}), while 
a 2D p-wave nanowire has their edge states located all along the 1D perimeter.

In the strong orbital limit, the phase transitions occur for critical values of the polar angle, following a simple analytical law that does not depend on sample details. With parallel ($x$) field orientation the transition law is also analytical, while for intermediate regimes the phase transitions are obtained numerically. We assess the consistency between the phase diagram and direct calculations of the Majorana modes in semi-infinite and finite 2D wires, 
emphasizing the importance of covariant grid discretizations for the latter.\cite{Governale,Janecek} 

The work is organized as follows. Section II introduces the physical system in detail while Sec.\ III contains the algorithm used 
to determine the phase-transition boundaries. 
In Sec.\ IV we discuss the phase diagrams for out-of-plane tiltings of the magnetic field with different strengths of the spin-orbit coupling. Sections V and VI deal with the actual Majorana solutions in semi-infinite and finite 2D Majorana nanowires, respectively. 
Section VII contains the conclusions of the work.

\section{Physical model}
\label{model}
The combination of s-wave superconductivity, Rashba interaction and an external magnetic field is a well known source of Majorana Fermions.\cite{Lutchyn}  
We consider electronic motion restricted to the $\hat{x}$ (longitudinal) and $\hat{y}$ (transverse) directions
in presence of these three effects. The homogeneous magnetic field points in an arbitrary direction and the edges are modeled as infinite square well potentials in the longitudinal and transverse directions (Fig.\ \ref{F1}). 

The nanowire physics is described by a Hamiltonian of the Bogoliubov-deGennes kind, split in the following way
\begin{equation}
\label{E1}
\mathcal{H}_{\it BdG} 
= 
\mathcal{H}_0
+
\mathcal{H}_Z
+
\mathcal{H}_R
+
\mathcal{H}_{\it orb}\; .
\end{equation}
The successive contributions to Eq.\ (\ref{E1}) are 
the zero-field and superconducting energies. 
\begin{equation}
\label{E1_0}
	\mathcal{H}_{0} = 
	\left( \frac{p_{x}^{2}+p_y^2}{2m} + V(x,y) -\mu \right)\tau_z 	+ \Delta_s\, \tau_{x}\; ;
\end{equation}
the Zeeman term
\begin{equation}
\label{E1_Z}
	\mathcal{H}_{\it Z} = 
\Delta_B\, \left(\sin\theta \cos\phi\, \sigma_x	+ \sin\theta \sin\phi\, \sigma_y +\cos\theta\, \sigma_z \right ) \, ;
\end{equation}
the Rashba coupling term
\begin{equation}
\label{E1_R}
	\mathcal{H}_{\it R} = 
\frac{\alpha}{\hbar}\, \left(\, p_x \sigma_{y} - p_y \sigma_{x}\, \right)\tau_z \, ,
\end{equation}
and, finally, the magnetic orbital terms 
\begin{equation}
\label{E1_orb}
	\mathcal{H}_{\it orb} = 
	\frac{\hbar^2}{2 m l^4_z}y^2 \tau_z - \frac{\hbar}{m l^2_z}y p_x - \frac{\alpha}{l^2_z}y \sigma_y \;.
\end{equation}
In Eqs.\ (\ref{E1_0})-(\ref{E1_orb})
we used the following Nambu-spinor convention, relating discrete components with spin $(\uparrow\downarrow)$
and isospin $(\Uparrow\Downarrow)$ as
$\Psi\equiv(
\Psi_{\uparrow\Uparrow},\Psi_{\downarrow\Uparrow},\Psi_{\downarrow\Downarrow},-\Psi_{\uparrow\Downarrow })^{T}$.

The contributions in Eq.\ (\ref{E1_0}) are, in left to right order, the kinetic, electrical potential $V$, chemical potential $\mu$ and superconducting $\Delta_s$ energies. The Pauli operators for isospin
(particle-hole) are represented by $\tau_{x,y,z}$ while those
for spin are $\sigma_{x,y,z}$.  The superconductor term represents an effective mean field approximation to more complicated interactions with a nearby s-wave superconductor. The Zeeman term, Eq.\ (\ref{E1_Z}), depends on parameter $\Delta_B$ and models the coupling of the spin 
with a magnetic field of arbitrary polar and azimuthal angles $(\theta,\phi)\equiv\hat{n}$. 

The Rashba coupling Eq.\ (\ref{E1_R}) is the result of the self-interaction between the quasiparticle spin with its own motion. This interaction is due to the presence of a transverse electric field representing an internal asymmetry in the confinement along $z$ that may be either intrinsic or externally induced. The first Rashba contribution, depending on $p_x \sigma_{y}$ is called the 1D Rashba term while the second one, $p_y \sigma_{x}$, is the Rashba mixing term. 

The joint effects of superconductivity, Zeeman and 1D Rashba terms give rise to independent Majorana states, one from each transverse band like in the 1D model. Each one of the modes 
has a different critical magnetic field 
$\Delta_{B,n}^{(c)}=\left[ \left(\mu -\epsilon_{n} \right)^2+\Delta^2_s \right]^{1/2}$,
with $n=1,2,\dots$, and $\epsilon_n$ the transverse mode energies.
Adding the Rashba mixing term to this scenario
changes the critical fields due to the 
coupling between different transverse bands. 
It effectively allows only one Majorana zero mode in parameter regions where the 1D Rashba term would yield an odd number of them (even-odd effect).\cite{Potter2,Serra}
This is further discussed in the Appendix as a particular analytical limit of the most general case presented below.
Besides, in 1D wires the effect of tilting the magnetic field implies the additional 
requirement of the so-called projection rule $\Delta_B\sin\theta\sin\phi<\Delta_s$.\cite{Osca3,Rex}
This is due
to the indirect gap closing of the infinite wire energy bands at $\pm k_c$
due to the tilted field, where $k_c$
is a non vanishing wave number. As discussed in Sec.\ IV, 
in 2D nanowires we find strong modifications of the critical magnetic fields, but
the projection rule still applies.

In a planar nanowire the perpendicular component of the magnetic field induces orbital motions of the nanowire quasiparticles. The magnetic 
orbital terms, Eq.\ (\ref{E1_orb}), describe this motion and their effect on the Majorana states is the central point of this article. These terms depend on the magnetic length $l_z$, defined as  $l_z^2=\hbar c/eB_z$, and they stem from the kinetic and Rashba energies with the magnetic substitution  $p_x \to p_x -\hbar y/l_z^2$ and adding the required Pauli matrix $\tau_z$ for proper particle-hole  symmetry.
In Eq.\ (\ref{E1_orb}) we assumed the Landau gauge centered on $y_c=0$, although our results are independent on this choice as discussed below.


All parameters of the complete Hamiltonian are constant inside the nanowire, modeled as a perfectly confining box with $L_x\gg L_y$. The numerical results of this work are presented in characteristic units of the problem obtained by taking  $\hbar$, $m$ and the width of the nanowire $L_y$  as reference values. That is, our length and energy units are, respectively, 
$L_{\it U}\equiv L_{y}$ and 
$E_{\it U}\equiv {\hbar^2}/{m L_{y}^2}$.
A spin-orbit length $L_{so}$ is usually defined as $L_{so}=\hbar^2/m\alpha$
but, as explained below, the Hamiltonian orbital terms will introduce an effective transversal confinement.
Therefore, the comparison between its characteristic length and the one of the nanowire 
real confinement is relevant. 
We notice that in our 
convention, the numerical value of $\alpha$ is precisely the ratio of transverse and spin-orbit lengths,
$\alpha/E_UL_U = L_y/L_{so}$.

\section{The matching method}
\label{secM}
In topological systems it is in general possible to relate the states of the bulk with those at 
the boundaries, a consequence of the bulk-to-edge correspondence principle. In our particular case this means that the semi-infinite Majorana solution $\Psi$ will be expressed as a linear superposition
of the infinite-nanowire eigensolutions  $\Psi^{(k)}$ (i.e., for the same Hamiltonian but disregarding the left and right edges),
\begin{equation}
	\Psi(x,y,\eta_\sigma,\eta_\tau) =\sum_{k}{ C_k\, \Psi^{(k)}(x,y,\eta_\sigma,\eta_\tau) }\;,
	\label{E3}
\end{equation}  
where spin and isospin variables are indicated with $\eta_\sigma$ and $\eta_\tau$, respectively.
Notice that the infinite-nanowire solutions $\Psi^{(k)}$ are characterized by a wave number $k$ that,
accounting for evanescent waves, may be a complex quantity.
In a given spin-isospin basis, $\chi_{s_\sigma}(\eta_\sigma)$ and  $\chi_{s_\tau}(\eta_\tau)$,
with $s_\sigma=\pm$ and $s_\tau=\pm$,
the infinite-wire solutions read
\begin{equation}
\Psi^{(k)}(x,y,\eta_\sigma,\eta_\tau) =
\sum_{s_\sigma s_\tau} {
\Psi_{ s_\sigma s_\tau}^{(k)}(y)\, e^{ikx}\,
\chi_{s_\sigma}(\eta_{\sigma})\,
\chi_{s_\tau}(\eta_{\tau})
}\; ,
\end{equation}  
where $\Psi^{(k)}_{s_\sigma s_\tau}(y)$ is a 1D 4-component spinor characteristic of the infinite-wire
solution with wave number $k$.
 
It has been demonstrated that a Majorana phase transition occurs in a semi-infinite nanowire when the propagating bands for the corresponding infinite nanowire perform a gap closing and reopening 
when increasing the magnetic field,
at 
vanishing energy and
wave number.\cite{Oreg,Osca3} 
Therefore, to determine the phase boundaries we only need to investigate the band structure at 
$k=0$ and $E=0$.
However, a full determination of the band spectrum for every set of Hamiltonian parameters by diagonalization is time consuming and computationally ineffective.
In accordance with this,  
it has been pointed out that in spite of the non locality of the topological states 
a full knowledge of the band spectrum 
is not necessary in general 
to determine the phase of a topological system, but the only relevant regions are those near the Dirac cones that appear at the phase transitions.\cite{Akhmerov2}
In our case, this implies searching the solutions of
\begin{equation}
\sum_{{s'}_\sigma {s'}_\tau}
\langle s_\sigma s_\tau | h | {s'}_\sigma {s'}_\tau\rangle\,
\Psi_{s'_\sigma s'_\tau}^{(0)}(y)=0\;,
	\label{E9}
\end{equation}
where $h$ is obtained neglecting all $p_x$-dependent terms in Eq.\ (\ref{E1}),
\begin{eqnarray}
	h &=& 
	\left( \frac{p_{y}^{2}}{2m} + V(y) -\mu \right)\tau_z 
	+ \Delta_s\, \tau_{x} - \frac{\alpha}{h}\, p_y \sigma_{x}\tau_z	
	\nonumber\\
	&&\!\!\!\!\!\!\!\!\!\! 
	+ \Delta_B\, \hat{\sigma}\cdot \hat{n} + \frac{\hbar^2}{2 m l^4_z}y^2 \tau_z - \frac{\alpha}{l^2_z}y \sigma_y \, .
	\label{E8}
\end{eqnarray}
Notice that with Eq.\ (\ref{E9}) we achieved a reduction to an effective 1D problem.

We can use the algorithm devised in Refs.\ \onlinecite{Serra,Osca1} to study the solutions
of Eq.\ (\ref{E9}). The particular parameter sets allowing such a solution will 
signal the gap closing of the 
original 2D Hamiltonian of Eq.\ (\ref{E1})
and thus the phase transition we are looking for.
The algorithm consists in solving Eq.\  (\ref{E9}) in a 1D grid as a linear system, assuming vanishing boundary conditions at $y=0$ and $y=L_y$. However, due to the homogeneous character of this linear system the trivial solution $\Psi_{s_\sigma s_\tau}^{(0)}(y)=0$ is always possible.
The algorithm discards the trivial solution 
by means of a matching point $y_m$,  where for an arbitrary pair of components 
$(s_\sigma s_\tau)=(st)$ a non vanishing wave function is imposed. In addition, 
continuity of the first derivative 
at the matching point is also imposed for the components other than 
$(st)$, 
\begin{eqnarray} 
\Psi^{(0)}_{st}(y_m) &=& 1\; , \label{E9b}\\
\!\!\!\!\!\!\!\!\!\!\!
\left(\frac{d^{(U)}}{dy}-\frac{d^{(L)}}{dy} \right) \Psi^{(0)}_{s_\sigma s_\tau}(y_m)&=&0\;,  \quad (s_{\sigma},s_{\tau})\neq (s,t)\,,
			\label{E10b}
\end{eqnarray}
where $d^{(U,L)}/dy$ denote grid derivatives using only upper (U) or lower (L) $y$-grid neighbors. 

Equations (\ref{E9b}) and (\ref{E10b}) are used at $y_m$ in place of the Bogoliubov-deGennes ones. 
In particular, Eq.\ (\ref{E9b}) makes the system no longer homogenous and such that it always admits a solution. The algorithm does not ensure the first-derivative continuity for the component $(st)$ at the matching point. 
Therefore, this condition is used to distinguish the physical from the non-physical solutions with the 
continuity {\em measure}
\begin{equation}
	\mathcal{F}=\left| \left(\frac{d^{(U)}}{dy}-\frac{d^{(L)}}{dy}\right)
	\Psi^{(0)}_{st}(y_m) \right|^2\,.
	\label{E11}
\end{equation}
As mentioned, the  ${\cal F}$ zeroes will
signal the desired gap closing boundaries of Eq.\ (\ref{E1}). 
Further details about the algorithm can be found in Ref.\ \onlinecite{Osca1}.

\section{Phase diagrams}
\label{PD}
\begin{figure}[t]
\centering
\resizebox{0.3\textwidth}{!}{
	\includegraphics{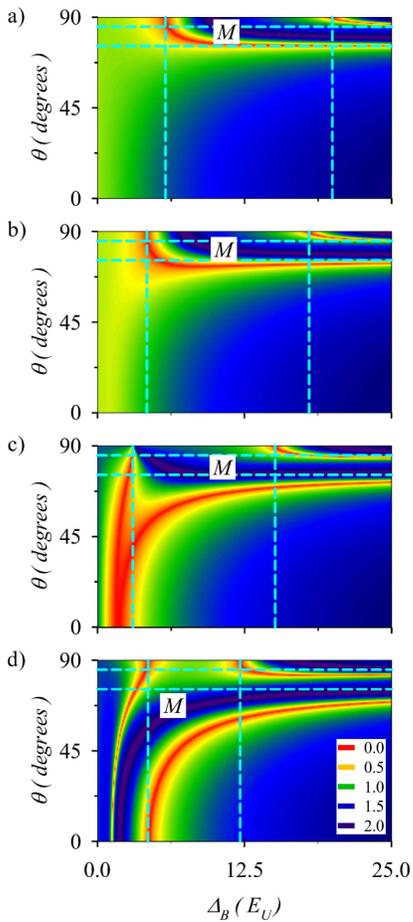}
}
\caption{$\mathcal{F}$ measure in a color (gray) scale as a function of  $\Delta_B$ and polar angle $\theta$. The azimuthal angle remains $\phi=0$. 
Zero values of $\mathcal{F}$ (red) signal the phase transition boundaries. Phases with a Majorana mode are labeled with an M. From top to bottom the panels correspond to $\alpha=0.1\, E_U L_U$ (a), $2\, E_U L_U$ (b), $\pi\, E_U L_U$ (c), $4\, E_U L_U$ (d). Dashed lines indicate the analytical limits. We have assumed $\Delta_s=3\, E_U$ and typical InAs parameters $g=15$, $m^*=0.033$. Panel (b) corrresponds to an InAs nanowire with 
$\alpha= 30\, {\rm meV}{\rm nm}$, $\Delta_s=0.3\, {\rm  meV}$ and $L_y=150\, {\rm nm}$ in a magnetic field range from zero to 6 T.}
\label{F2}
\end{figure}

Figure \ref{F2}b shows the phase transition boundaries obtained with the matching method for an $L_y=150\; {\rm nm}$ nanowire with material parameters typical of InAs and a magnetic field strength between $0$ and $6$ T ($\Delta_B=0-25\, E_U$). Phase boundaries are signaled by zero values of ${\cal F}$ (red lines). Only $ {\cal F}=0$  values represented by red lines have physical sense and  
the color scale is only indicating the measure deviation from zero. 
As mentioned, the energy unit scales as $L_y^{-2}$, such that in a  300 nm wire  $25 E_U$ would correspond to 1.5 T. The other panels of Fig.\ \ref{F2} correspond to lower (\ref{F2}a) and higher (\ref{F2}c,d) values of the Rashba coupling strength. We will explicitly calculate the zero modes for particular sets of parameters below. Here, let us anticipate that the orbital terms do not change the even-odd effect of multiband nanowires.\cite{Lutchyn2,Potter,Potter2,sanjose} 

The phases in Fig.\ \ref{F2} contain either no Majoranas or at most one Majorana mode in regions labeled with an M. The main result of this article is that orbital effects do not destroy Majoranas into other phases.
However, they do lead to characteristic phase maps where the Majorana states survive sizable vertical tiltings (even up to $\theta=0$ in Fig.\ \ref{F2}d). The change in transition boundaries is caused by two reasons. First, by the change in the effective transversal confinement due the first two  terms of Eq.\ (\ref{E1_orb}). Note that the first contribution leads to a harmonic confining. Second, the third term of Eq.\ (\ref{E1_orb}) can 
be understood as an effective inhomogeneous magnetic field pointing in $y$ direction due to the 
combination of Rashba and orbital effect.

It is possible to give analytical expressions of the phase boundaries in particular 
limits (see Appendix).
For $\theta=90^\circ$ and $\phi=0$ the critical magnetic fields read
\begin{equation}
	\Delta_{B,n}^{(c)}=\sqrt{\left(\mu -\epsilon_{n}+\frac{m\alpha^2}{2\hbar^2} \right)^2+\Delta^2_s}\,,
	\label{E20}
\end{equation}
where $n=1,2\dots$ and  $\epsilon_n$ are the (transverse) square well eigenenergies. This analytical result extends recent findings from other authors \cite{Lutchyn2,sanjose} who assumed that the contribution in parenthesis in Eq.\ (\ref{E20}) is an effective chemical potential from subband $n$, without specifying its detailed $\alpha$ dependence. Analogously, in the strong orbital limit $\Delta_B>>(\hbar^2/mL_y^2,m\alpha^2/\hbar^2,\Delta_s,\mu)$ the critical angles are 
\begin{equation}
	\theta^{(c)}_n=\arccos \left( \frac{gm^*}{4(n-\frac{1}{2})} \right) \,,
	\label{E21}
\end{equation}
where $g$ is the gyromagnetic factor and $m^*=m/m_e$ the ratio between the electron effective and bare masses ($m$ and $m_e$ respectively). In this limit quasiparticles are confined by the effective $\Delta_B$-dependent harmonic potential caused by the first and second terms of Eq.\ (\ref{E1_orb}), independently of the real nanowire transversal boundaries. Quite remarkably, Eq.\ (\ref{E21}) is independent of the magnetic field, Rashba and superconductivity strengths, as well as on the specific wire width $L_y$. In this sense, the critical angles are rather universal. The values of Eqs.\ (\ref{E20}) and (\ref{E21}) for the particular parameters used in Fig.\ \ref{F2} are overprinted as vertical and horizontal dashed lines, respectively. As shown, the numerical values match the analytical ones in their corresponding limits. In Figs.\ \ref{F2}a,b the transitions boundaries do not deviate substantially from the analytical laws in all the plot, while Figs.\ \ref{F2}c,d show large differences for intermediate values of the parameters. 
 
The structure of the phase transitions in Figure \ref{F2}a is typical for cases when the kinetic orbital effects already dominate around the first transition boundary with increasing $\Delta_B$. The phase boundaries just bend from a vertical to a horizontal line due to the 
effective change of the transversal confinement in the nanowire, from square well to a $\Delta_B$-dependent harmonic confinement [Eq.\ (\ref{E1_orb})]. Assuming $\mu\approx 0$, the conditions for this simpler phase diagrams  ($l_z$ shortest scale) can be written as a triple inequality

\begin{equation}
\label{ineq}
\Delta_{B,1}^{(c)}\cos\theta_1^{(c)}
\gg
\frac{gm^*}{4}\left(E_U, \frac{\alpha^2}{E_U L_U^2}, \Delta_s\right)\; .
\end{equation}

Figure \ref{F2}b (InAs with $L_y= 150\, {\rm nm}$)  still presents a phase diagram qualitatively similar to Fig.\ \ref{F2}a although the second
inequality of Eq.\ (\ref{ineq}) is not well satisfied (see Tab.\  \ref{table1} in Appendix). 
On the other hand, Figures \ref{F2}c and d show the modifications of the phase diagram as $\alpha$ increases in effective units. As anticipated, the deviations can even allow a Majorana state in perpendicular field
($\theta=0$) in Fig.\ \ref{F2}d. However, so large spin orbit strengths give rise to complicated phase maps that strongly deviate from the analytical limits. 

The strong-$\alpha$ effects seen in the lower panels of Fig.\ \ref{F2} are caused by
the term $-\alpha y\sigma_y/l_z^2$ of Eq.\ (\ref{E1_orb}). Indeed, this term
effectively adds a component along $y$ to the magnetic field. Therefore, 
the effective angle $\theta_e$ is such that $\theta_e>\theta$, thus explaining the downwards shift of the lower phase transition boundary in Figs.\ \ref{F2}c and d. Figures \ref{F2}a,b do not change
with other azimuthal angles while in the strong-$\alpha$ diagrams (Fig.\ \ref{F2}c,d) $\phi$ 
modifies the precise boundary positions for $\theta=90^\circ$ (vertical lines),
but not the horizontal asymptotes and the overall qualitative behavior. Note that,
as mentioned in Sec.\ \ref{model}, with $\phi\ne 0$ there is 
an additional requirement for the existence of Majorana modes, the projection 
rule $\Delta_B \sin\theta \sin\phi<\Delta_s$.\cite{Osca3,Rex} 
However,
the effective tilting of the magnetic field towards $y$ caused by the term $-\alpha y\sigma_y/l_z^2$  
does not modify the projection rule because this effective tilting is in opposite directions for 
positive and negative $y$'s while the projection rule applies only to homogeneous magnetic fields.

\section{Semi-infinite nanowires}
Explicit zero-energy eigenstates can be obtained in semi-infinite and finite nanowires in order to confirm the above 
phase diagrams. We have checked that either one or none Majoranas are obtained in the corresponding regions of Fig.\ \ref{F2}. 
In the semi-infinite system exact zero-energy solutions can be studied with the complex band structure method of Refs.\ \onlinecite{Serra,Osca3}. 
The calculation is carried out solving the boundary condition
\begin{equation}
	\sum_k C_k \Psi^{(k)}_{s_\sigma,s_\tau}(y)=0 \,,
	\label{E22}
\end{equation}
where the allowed complex wave numbers $\{ k \}$ and the $\Psi^{(k)}_{s_\sigma s_\tau}(y)$ functions are 
obtained with the matching method  discussed previously in Sec.\ \ref{secM}, 
with the only difference that $k$ now is not vanishing.

Equation (\ref{E22}) can be reworked into
\begin{equation}
	\sum_k \mathcal{M}_{k'k}C_k=0 \,,
	\label{E23}
\end{equation}
with the matrix
\begin{equation}
	\mathcal{M}_{k'k}=\sum_{s_\sigma,s_\tau} \int dy\,  \Psi^{(k')*}_{s_\sigma,s_\tau}(y) \Psi^{(k)}_{s_\sigma,s_\tau}(y)\,.
	\label{E24}
\end{equation}

\begin{figure}[t]
\centering
\resizebox{0.4\textwidth}{!}{
	\includegraphics{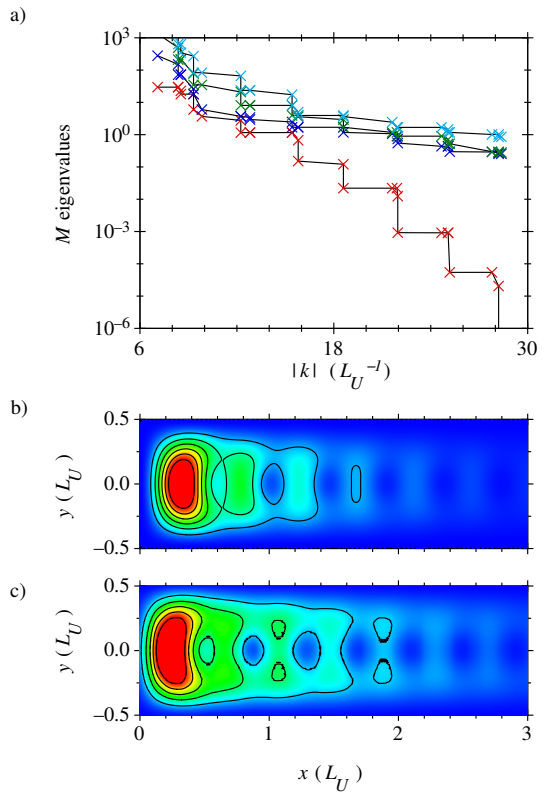}
}
\caption{a) Evolution of the lower eigenvalues of matrix $\mathcal{M}$ when increasing
the number of evanescent modes, as given by a cutoff $|k|$. 
b) Majorana density function in a semi-infinite nanowire for the null eigenvector of ${\cal M}$. 
c) Majorana density function in a finite nanowire with $L_x=20 L_U$ calculated by direct diagonalization of the Hamiltonian using covariant derivative discretization. 
The  three panels correspond to $\Delta_B=10\, E_U$, $\theta=75^\circ$ and the rest of
parameters as in Fig.\ \ref{F2}c.}
\label{F3}
\end{figure}

Equation (\ref{E23}) shows that, when enough wave numbers are included, each Majorana state is represented by a null-space eigenvector of matrix  $\mathcal{M}$.  In Fig.\ \ref{F3}a we can see the convergence of the ${\cal M}$ eigenvalues with the  cut off in wave number for a particular point of Fig.\ \ref{F2}c. Clearly, the lower eigenvalue vanishes asymptotically indicating  that for this point of the phase diagram a Majorana mode is present as expected. In Fig. \ref{F3}b we can see the corresponding density function, confirming the edge character of the mode, as also expected for a Majorana.

\subsection{Decay lengths}
\label{dcay}

\begin{figure}[t]
\centering
\resizebox{0.4\textwidth}{!}{
	\includegraphics{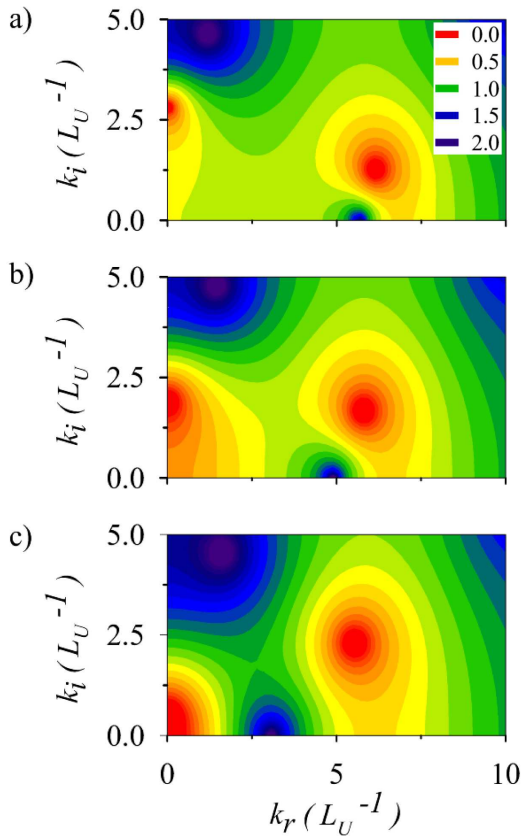}
}
\caption{ $\mathcal{F}$ measure in a color (gray) scale showing the position of the allowed wave numbers as zeroes (red islands).
The parameters used are $g=15$, $m^*=0.033$, $\alpha=\pi\, E_U L_U$, $\Delta_s=3\, E_U$, 
$\Delta_B=10\, E_U$. Panels from top to bottom are for different polar angles: $\theta=68^\circ$ (a),   $\theta=67^\circ$ (b),  $\theta=66^\circ$ (c).}
\label{F5}
\end{figure}

Within our complex-band-structure approach to the semi-infinite nanowire we can estimate the length of the Majorana decay tail from the imaginary part of the allowed wave numbers.  The lower the imaginary part, the longer the Majorana decay tail (and thus the required length of the nanowire to contain it without distortion). Figure \ref{F5} shows a typical evolution of the wave numbers (red islands) in the complex plane as the polar angle is approaching the critical value. In the sequence from  upper to lower panels, one of the wave numbers moves along the imaginary 
axis towards the origin; the phase transition being signaled by one mode touching the origin (lower panel).
 
We calculate the required nanowire length with the smallest imaginary wave number  of the set of all allowed wave numbers $\{k^{(m)}\}$. However, as shown in Fig.\ \ref{F5} the smallest imaginary part $\Im(k^{(m)})\equiv k^{(m)}_i$ changes from an approximately  fixed mode to the one touching the origin when approaching the phase transition. We define the mode  length $L_m$ as two times the length needed for the wave function to drop to one percent of its maximum, that is  $e^{-k_i^{(m)}L_m/2}=0.01$. An estimate of the nanowire length for undistorted Majoranas is simply the maximum of all 
mode lengths.

\begin{figure}[t]
\centering
\resizebox{0.35\textwidth}{!}{
	\includegraphics{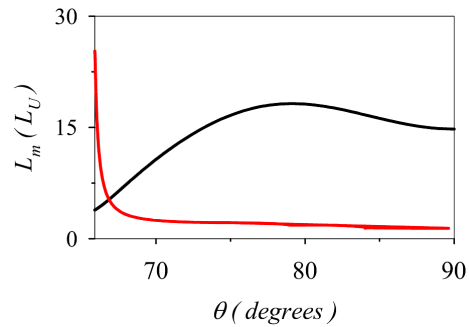}
}
\caption{Mode lengths $L_m$  (defined in Sec.\ \ref{dcay}) for the two wave numbers shown in Fig.\ \ref{F5}. Note that the required nanowire length at each $\theta$  is the higher of both curves.}
\label{F6}
\end{figure}

In Fig.\ \ref{F6} we show the mode lengths of the two allowed wave numbers of Fig.\ \ref{F5}. As we decrease the polar angle from 90 degrees, the needed nanowire length (the higher of the two curves) remains more or less stable until $\theta$ approaches the critical value. A few degrees before the transition the Majorana contracts before diverging to infinity at the phase transition angle.

\section{Finite nanowires}

The phase diagram can also be checked with full diagonalizations of nanowires with large, but finite, $L_x$. Though more realistic,  
this approach is conceptually more qualitative, since finite-nanowire Majoranas are not exact zero modes but small energy modes 
(the smaller the energy the larger $L_x$). Equivalently, the phase boundaries become blurred due to the finite size effect.
Figure \ref{F3}c shows the density of the finite nanowire Majorana corresponding to the semi infinite one of Fig.\ \ref{F3}b. Differences are small, 
just a slight distortion and a somewhat longer decay tail of the finite-nanowire density.

In the finite nanowire diagonalization with orbital terms we have found it crucial using a covariant grid discretization.\cite{Governale,Janecek}
Otherwise, numerical artificial biases wrongly suggest that Majoranas are always destroyed  by orbital terms,\cite{Lim} in clear contradiction 
with the phase diagram (Fig.\ \ref{F2}) and the semi-infinite wire analysis. In essence, the covariant discretization amounts to expressing the canonical momentum components as symmetry-like transformations. For instance, 
\begin{equation} 
\Pi_x   \equiv -i\hbar \frac{\partial}{\partial x} - \hbar \frac{ y}{l_z^2} 
= e^{iyx/l_z^2}\left( -i\hbar \frac{\partial}{\partial x}\right) e^{-iyx/l_z^2}  \,.
\label{E50}
\end{equation}
Although these two representations of the canonical operator are equivalent in the continuous limit, they are not on a discrete grid. 

As demonstrated in Ref.\ \onlinecite{Janecek} the covariant derivative preserves by construction the gauge invariance of the solutions while a non-covariant 
treatment only does that for extremely fine discretizations, unfeasible in our case. 
Changing the gauge origin usually constitutes a 
severe difficulty for numerical discretizations not using covariant derivative formulations.
In our case, we can introduce an arbitrary gauge center $y_c$ for the
canonical momentum, generalizing Eq.\ (\ref{E50}) to
\begin{eqnarray} 
\Pi_x&=& e^{i(y-s_\tau y_c)x/l_z^2}\left( -i\hbar \frac{\partial}{\partial x}\right) e^{-i(y-s_\tau y_c)x/l_z^2}  \,,
\end{eqnarray}
where
the isospin sign $s_\tau=\pm$ is introduced in order to preserve the particle-hole 
symmetry of the Bogoliubov-deGennes equation. We have checked that our numerical results for the finite nanowire diagonalization as, e.g., 
in the lower panel of Fig.\ \ref{F3}, do not depend on the choice of $y_c$, thus proving the gauge invariance of the finite system results. 
We have also obtained good agreement of the finite nanowire diagonalizations and the results of the semi-infinite system regarding the existence or absence of a zero mode in the different regions of the phase diagrams (Fig.\ \ref{F2}), again proving the reliability of the method. Notice that the semi-infinite solution, being purely 1D, can be obtained in very dense $y$ grids, while the finite system 2D diagonalization requires much coarser
$xy$ grids. 

\section{Conclusions}
In this work we have shown that the orbital motions caused by perpendicular components of the magnetic field in planar 2D
nanowires give rise to a rich phase diagram, with regions containing Majoranas for sizable vertical tilting of the magnetic field.
In fact, with proper parameters, it is possible to find  Majoranas in a fully perpendicular field. We have developed a general numerical method to obtain the Majorana phases in nanowires in a computer efficient way and we have checked this method against alternative calculations for semi-infinite and finite nanowires.  Analytical expressions of the transition boundaries in asymptotic regions have been found. 
For realistic parameter values (weak $\alpha$) these analytical expressions are a good approximation in general and not only asymptotically. 
In the strong orbital limit the critical angles are independent of sample details. Finally, the relevance of the covariant grid discretization 
for the finite nanowire diagonalization has been pointed out. 

\begin{acknowledgments}
We acknowledge useful discussions with R. L\'opez and D. S\'anchez.
This work was funded by MINECO-Spain (grant FIS2011-23526),
CAIB-Spain (Conselleria d'Educaci\'o, Cultura i Universitats) and 
FEDER. We hereby acknowledge the PhD grant provided by the University 
of the Balearic Islands.
\end{acknowledgments} 

\appendix*

\section{Analytical limits}
\subsection{Longitudinal magnetic field}

When the magnetic field is along $x$ (see axis orientations in Fig.\ 1) the phase transition law is fully analytical. 
As discussed in Sec.\ \ref{secM}, 
finding the phase transition implies searching for the zero energy eigenstates of the simplified Hamiltonian 
$h$ given in Eq.\ (\ref{E8}).
For an $x$-oriented field $\l_z^{-2}=0$ and all orbital terms vanish. Assuming also a vanishing bottom potential
for $V(y)$ we have 
\begin{equation}
	h = 
	\left( \frac{p_{y}^{2}}{2m} -\mu \right)\tau_z 
	+ \Delta_s\, \tau_{x} 
	+ \left( \Delta_B 
	- \frac{\alpha}{\hbar} p_y \tau_z\right) \sigma_x \;.
	\label{E30}
\end{equation}

The eigenstates of Eq.\ (\ref{E30}) can be obtained analytically noticing that the linear $p_y$ term from the Rashba interaction
can be absorbed in the kinetic term 
\begin{equation}
	h = 
	\left( \frac{\tilde{p}_{y}^{2}}{2m}-\frac{m\alpha^2}{2\hbar^2} -\mu \right)\tau_z 
	+ \Delta_s\, \tau_{x} 
	+ \Delta_B\sigma_x \;,
	\label{E30n}
\end{equation}
where $\tilde{p}_y=p_y-m\alpha\sigma_x/\hbar$. Using a basis of square-well eigenstates of energies
$\epsilon_n=\hbar^2\pi^2 n^2 /(2 m L_y^2)$, the matrix to diagonalize is
\begin{equation}
	h\equiv\left(
		\begin{array}{cc}			
			\epsilon_{n}-\mu -\frac{m\alpha^2}{2\hbar^2} & \Delta_s \\			
			\Delta_s & -(\epsilon_{n}-\mu -\frac{m\alpha^2}{2\hbar^2} )
 		\end{array}
  \right)\,.
  \label{E50b}
\end{equation}
The diagonalization of this simplified Hamiltonian yields the eigenenergies
\begin{equation}
	E_{n s_1 s_2}=s_{1}\Delta_B+s_{2}\sqrt{\left(\mu -\epsilon_{n}+\frac{m\alpha^2}{2\hbar^2}
	\right)^2+\Delta^2_s}\,,
	\label{E60}
\end{equation}
with $n=1,2,\dots$, $s_1=\pm1$ and $s_2=\pm1$.
Of the four eigenenergies only the two with opposite $s_1$ and $s_2$ can lead to a zero energy solution at the critical values
\begin{equation}
	\Delta_{B,n}^{(c)}=\sqrt{\left(\mu -\epsilon_{n}+\frac{m\alpha^2}{2\hbar^2} 
	\right)^2+\Delta^2_s}\,.
	\label{E70}
\end{equation}

\begin{figure}
\centering
\resizebox{0.45\textwidth}{!}{
	\includegraphics{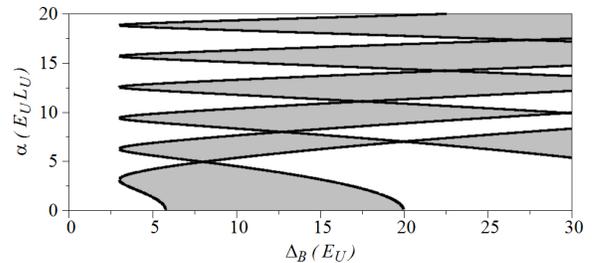}
}
\caption{ a) Phase transition boundaries for a 2D planar nanowire with a longitudinal ($x$) magnetic field.
We have assumed $\Delta_s=3 E_U$ and $\mu=0$. The shaded regions correspond to topological phases with a Majorana zero mode.}
\label{F4}
\end{figure}

Equation (\ref{E70}) with $n=1,2,\dots$ gives the critical Zeeman parameter of 
phase transitions for a two dimensional nanowire in parallel magnetic field. 
Notice that in Eq.\ (\ref{E70}) $n$ has to be interpreted simply as an ordering index of the successive 
transitions, and not as a label of independent transverse modes. These latter interpretation 
would be wrong, since different transverse modes are coupled through the Rashba mixing term and one can not
associate a particular transverse mode with a particular transition point.
Shaded regions in Fig.\  \ref{F4} contain one Majorana mode, while white regions have none.
There are no regions with multiple Majoranas due to the energy 
splittings induced by the Rashba mixing in planar nanowires.\cite{Potter2,Serra}

\subsection{Strong orbital limit}

When the kinetic orbital effect overcomes both the confinement by the transverse 
square well and the Rashba term,
the magnetic length $l_z\equiv \sqrt{\hbar c / e B_z}$
is smaller than $l_y$ (here we define $l_y=L_y$ of Fig.\ \ref{F1})
 and also smaller than the Rashba length $l_\alpha\equiv\hbar^2/m\alpha$.
In the limit $l_z\ll (l_y,l_\alpha)$ it is possible to derive an 
analytical expression of the transition boundaries.
Neglecting the square well $V(y)$ and the Rashba terms in Eq.\ (\ref{E8}) we find
\begin{equation}
	h = 
	\left( \frac{p_{y}^{2}}{2m} +  \frac{\hbar^2}{2 m l^4_z}y^2 -\mu \right)\tau_z 
	+ \Delta_B\, \vec\sigma\cdot\hat{n}
	+\Delta_s\, \tau_{x}\;.
	\label{E31}
\end{equation}

The eigenvalues of Eq.\ (\ref{E31}) are straightforward in a basis 
$|n s_\sigma s_\tau\rangle$, where $n=1,2,\dots$ represent now harmonic oscillator
eigenstates, $s_\sigma=\pm$ indicates spin eigenstates in 
direction $\hat{n}$, while $s_\tau=\pm$ indicates isospin in direction $z$.
Since the $h$ matrix is diagonal in spin, we can diagonalize each subspace
independently. For instance, the matrix for $s_\sigma=+$ reads
\begin{equation}
\label{hnew}
\left(
\begin{array}{cc}
\epsilon_n^{(ho)}-\mu+\Delta_B & \Delta_s \\
\Delta_s & -\left(\epsilon_n^{(ho)}-\mu\right)+\Delta_B
\end{array}
\right)\;,
\end{equation}
with $\epsilon_n^{(ho)}=(n-1/2)\hbar^2/ml_z^2$. The eigenvalues of
Eq.\ (\ref{hnew}) are easily found, as well as those of the 
analogous matrix for spin $s_\sigma=-$.

\begin{table}[t]
\begin{ruledtabular}
\begin{tabular}{ccccc}
 Panel & $\Delta_{B,1}^{(c)}\cos\theta_1^{(c)}$ & $\frac{gm^*}{4 E_U}$ &
 $\frac{gm^*\alpha^2}{4 E_U^2L_U^2}$ \rule[-0.3cm]{0cm}{0.1cm} & 
 $\frac{gm^*\Delta_s}{4}$ \\
 \hline
 a) & 1.43 & 0.12 & 0.001 & 0.36 \\
 b) & 1.03 & 0.12 & 0.48 & 0.36 \\
 c) & 0.74 & 0.12 & 1.22 & 0.36 \\
 d) & 1.06 & 0.12 & 1.98 & 0.36
\end{tabular}
\end{ruledtabular}
\caption{Numerical values in effective units of the inequalities, Eq.\ (11) of the paper, corresponding to 
the four panels in Fig.\ 2 of the paper.}
\label{table1}
\end{table}

The null-eigenvalue condition for $h$ is now
\begin{equation}
\label{E8n}
\Delta_B = \sqrt{\left[(n-1/2)\frac{\hbar^2}{ml_z^2}-\mu\right]^2+\Delta_s^2}\; ,
\end{equation}
that looks similar to Eq.\ (\ref{E70}). An essential difference, however, 
is that the r.h.s.\ in Eq.\ (\ref{E8n}) depends itself on the 
Zeeman parameter $\Delta_B$ through $l_z$. It is
\begin{equation}
\frac{\hbar^2}{ml_z^2}=\frac{4}{gm^*}\Delta_B \cos\theta\; ,
\end{equation}
where $m^*$ is the ratio of effective to bare mass, $m=m^* m_e$, while $g$ is the 
gyromagnetic factor defined from the Zeeman parameter by $\Delta_B \equiv g \mu_B B /2$.
From Eq.\ (\ref{E8n}) we finally arrive at the following relation 
\begin{equation}
\cos\theta = 
\frac{gm^*}{4}
\frac{
\sqrt{\Delta_B^2-\Delta_s^2}+\mu
}
{
\left(n-\frac{1}{2}\right)\Delta_{B}
}\; .
\end{equation}
For large enough $\Delta_B$, as compared to $\Delta_s$ and $\mu$, this leads to the 
prediction of field-independent critical angles
\begin{equation}
\label{ES11}
\cos\theta^{(c)}_n = 
\frac{gm^*}{4 
\left(n-\frac{1}{2}\right)
}\; ,
\end{equation}
as given in Eq.\ (\ref{E21}).

The triple inequality $\l_z\ll (l_y,l_\alpha,\l_s)$, where we define $l_s\equiv\sqrt{\hbar^2/m\Delta_s}$, leads,
when written in effective units, to Eq.\  (\ref{ineq}). In this situation the 
phase diagram does not deviate much from the straight lines of the 
analytical limits,  Eqs.\ (\ref{E70}) and (\ref{ES11}).
Table \ref{table1} contains the numerical 
values of the inequality sides for the four panels in Fig.\ 2. 
While panel a) fulfills all conditions,
for the rest of panels the second inequality degrades as $\alpha$ increases from panel b) to d).
This explains the deviations in those panels from the analytical limits.

\bibliographystyle{apsrev4-1}
\bibliography{Articulo_4L}

\end{document}